\documentclass[sts]{imsart}
\usepackage{amsthm,amsmath,amssymb,amsthm}
\usepackage{natbib}
\usepackage[colorlinks,citecolor=blue,urlcolor=blue]{hyperref}

\usepackage{commath}

\usepackage{tikz}
\usetikzlibrary{arrows}

\startlocaldefs
\DeclareMathOperator{\E}{\mathbb{E}}
\DeclareMathOperator{\ind}{\mathbb{I}}
\newcommand{\union}{\bigcup}

\theoremstyle{definition}
\newtheorem{algorithm}{Algorithm}
\endlocaldefs

\begin{document}

\begin{frontmatter}
  \title{A Review of Self-Exciting Spatio-Temporal Point Processes and Their
    Applications}
  \runtitle{Self-exciting spatio-temporal point processes}

  \author{\fnms{Alex}
    \snm{Reinhart}\corref{}\ead[label=e1]{areinhar@stat.cmu.edu}}
  \address{Department of Statistics \& Data Science, Carnegie Mellon University,
    5000 Forbes Ave, Pittsburgh, PA 15213, USA \printead{e1}.}

  \runauthor{A.~Reinhart}

  \date{\today}

  \begin{abstract}
    Self-exciting spatio-temporal point process models predict the rate of
    events as a function of space, time, and the previous history of events.
    These models naturally capture triggering and clustering behavior, and have
    been widely used in fields where spatio-temporal clustering of events is
    observed, such as earthquake modeling, infectious disease, and crime. In the
    past several decades, advances have been made in estimation, inference,
    simulation, and diagnostic tools for self-exciting point process models. In
    this review, I describe the basic theory, survey related estimation and
    inference techniques from each field, highlight several key applications,
    and suggest directions for future research.
  \end{abstract}

\end{frontmatter}

\section{Introduction}

Self-exciting spatio-temporal point processes, an extension of temporal Hawkes
processes, model events whose rate depends on the past history of the process.
These have proven useful in a wide range of fields: seismological models of
earthquakes and aftershocks, criminological models of the dynamics of crime,
epidemiological forecasting of the incidence of disease, and many others. In
each field, the spatio-temporal distribution of events is of scientific and
practical interest, both for prediction of new events and to improve
understanding of the process generating the events. We may have a range of
statistical questions about the process: does the rate of events vary in space
and time? What spatial or temporal covariates may be related to the rate of
events? Do events trigger other events, and if so, how are the triggered events
distributed in space and time?

Regression is a natural first approach to answer these questions. By dividing
space into cells, either on a grid or following natural or political boundaries,
and dividing the observed time window into short discrete intervals, we can
aggregate events and regress the number of events observed in a given cell and
interval against spatial and temporal covariates, prior counts of events in
neighboring cells, and so on. This approach has been widely used in
applications. However, it suffers several disadvantages: most notably, the
Modifiable Areal Unit Problem means that estimated regression coefficients and
their variances may vary widely depending on the boundaries or grids chosen for
aggregation, and there is no natural ``correct'' choice
\citep{Fotheringham_1991}.

Instead, we can model the rate of occurrence of events directly, without
aggregation, by treating the data as arising from a point process. If the
questions of scientific interest are purely spatial, the events can be analyzed
using methods for spatial point processes \citep{Diggle:2014}, and their times
can be ignored. If time is important, descriptive statistics for the first- and
second-order properties of a point process, such as the average intensity and
clustering behavior, can also be extended to spatio-temporal point processes
\citep[chapter 11]{Diggle:2014}.

When descriptive statistics are not enough to understand the full dynamics of
the point process, we can use spatio-temporal point process models. These models
estimate an intensity function which predicts the rate of events at any spatial
location \(s\) and time \(t\). The simplest case is the homogeneous Poisson
process, where the intensity is constant in space and time. An example of a more
flexible inhomogeneous model is the log-Gaussian Cox process, reviewed by
\citet{Diggle:2013iv}, in which the log intensity is assumed to be drawn from a
Gaussian process. With a suitable choice of spatio-temporal correlation
function, the underlying Gaussian process can be estimated, though this can be
computationally challenging.

Cluster processes, which directly model clustering behavior, split the process
in two: cluster centers, generally unobserved, are drawn from a parent process,
and each cluster center begets an offspring process centered at the parent
\citep[Section~6.3]{Daley:2003v1}. The observed process is the superposition of
the offspring processes. A common case is the Poisson cluster process, in which
cluster centers are drawn from a Poisson process; special cases include the
Neyman--Scott process, in which offspring are also drawn from a Poisson process,
and the Mat\'{e}rn cluster process, in which offspring are drawn uniformly from
disks centered at the cluster centers. Common cluster processes, other
spatio-temporal models, and descriptive statistics were reviewed by
\citet{Gonzalez:2016fl}.

In this review, I will focus on \emph{self-exciting} spatio-temporal point
process models, where the rate of events at time \(t\) may depend on the history
of events at times preceding \(t\), allowing events to trigger new events. These
models are characterized by a \emph{conditional} intensity function, discussed
in Section~\ref{pp-theory}, which is conditioned on the past history of the
process, and has a direct representation as a form of cluster process.
Parametrization by the conditional intensity function has allowed a wide range
of self-exciting models incorporating features like seasonality, spatial and
temporal covariates, and inhomogeneous background event rates to be developed
across a range of application areas.

Dependence on the past history of the process is not captured by log-Gaussian
Cox processes or spatial regression, but can be of great interest in some
applications: the greatest development of self-exciting models has been in
seismology, where prediction of aftershocks triggered by large earthquakes is
important for forecasting and early warning. However, literature on theory,
estimation, and inference for self-exciting models has largely been isolated
within each application, so the purpose of this review is to synthesize these
developments and place them in context, drawing connections between each
application and paving the way for new uses.

Self-exciting models can be estimated using standard maximum likelihood
approaches, discussed in Section~\ref{maximum-likelihood} below. Once a
self-exciting model is estimated, we are able to answer a range of
scientifically interesting questions about the dynamics of their generating
processes. Section~\ref{declustering} reviews \textit{stochastic declustering}
methods, which attribute events to the prior events which triggered them, or to
the underlying background process, using the estimated form of the triggering
function. Section~\ref{simulation} then introduces algorithms to efficiently
simulate new data, and Section~\ref{asymptotic-normality} discusses methods for
estimating model standard errors and confidence intervals. Bayesian approaches
are discussed in Section~\ref{bayesian}, and general model-selection and
diagnostic techniques in Section~\ref{model-selection}.

Finally, Section~\ref{applications} introduces three major application areas of
self-exciting spatio-temporal point processes: earthquake forecasting, models of
the dynamics of crime, and models of infectious disease. These demonstrate the
utility of self-exciting models and illustrate each of the techniques described
in Section~\ref{estimation-inference}. Section~\ref{social-networks} introduces
a further extension of self-exciting point processes, extending them from
spatio-temporal settings to applications involving events occurring on networks.

\section{Self-Exciting Spatio-Temporal Point Processes}
\label{pp-theory}

\subsection{Hawkes Processes}
\label{hawkes}


Consider a temporal simple point process of event times \(t_i \in [0, T)\), such
that \(t_i < t_{i + 1}\), and a right-continuous counting measure \(N(A)\),
defined as the number of events occurring at times \(t \in A\). Associated
with the process is the history \({\cal H}_t\) of all events up to time \(t\).
We may characterize the process by its \textit{conditional intensity}, defined
as
\[
  \lambda(t \mid {\cal H}_t) = \lim_{\Delta t \to 0} \frac{\E\left[N\left([t, t
        + \Delta t)\right) \mid {\cal H}_t\right]}{\Delta t}.
\]

The self-exciting point process model was introduced for temporal point
processes by \citet{Hawkes:1971}. Self-exciting processes can be defined in
terms of a conditional intensity function in the equivalent forms
\begin{align*}
  \lambda(t \mid {\cal H}_t) &= \nu + \int_0^t g(t - u) \dif N(u)\\
                             &= \nu + \sum_{i: t_i < t} g(t - t_i),
\end{align*}
where \(\nu\) is a constant background rate of events and \(g\) is the
triggering function which determines the form of the self-excitation. The
process is called ``self-exciting'' because the current conditional intensity is
determined by the past history \({\cal H}_t\) of the process. Depending on the
form chosen for the triggering function \(g\), the process may depend only on
the recent history (if \(g\) decays rapidly) or may have longer term effects.
Typically, because \(\lambda(t \mid {\cal H}_t) \geq 0\), we require \(g(u) \geq
0\) for \(u \geq 0\) and \(g(u) = 0\) for \(u < 0\).

\begin{figure*}
  \begin{tikzpicture}
    [>=latex,
    event/.style={black,circle,fill,inner sep=1.5pt},
    triggered1/.style={black,circle,draw,inner sep=1.5pt,fill=white},
    triggered2/.style={black,shape=rectangle,draw,inner sep=2.5pt,fill=white}]

    \begin{scope}[shift={(0, 0.5)}]
      \draw[->] (0, 0) node[left] {Observed} -- (11, 0) node[right] {$t$};
      \foreach \x in {0, 2, 4, 6, 8, 10}
      \draw[shift={(\x, 0)},color=black] (0pt, 3pt) -- (0pt, -3pt);

      \foreach \x in {1.2, 1.8, 2.6, 3.2, 3.4, 4.3, 5.1, 5.4, 5.9, 6.7, 8.4, 9.5}
      \draw[shift={(\x, 0)}] (0,0) node[event] {};
    \end{scope}

    \begin{scope}[shift={(0, -1)}]
      \draw[->] (0, 0) node[left] {Background} -- (11, 0);
      \foreach \x in {0, 2, 4, 6, 8, 10}
      \draw[shift={(\x, 0)},color=black] (0pt, 3pt) -- (0pt, -3pt);

      \node (bg1) at (1.2, 0) [event] {};
      \node (bg2) at (3.2, 0) [event] {};
      \node (bg3) at (5.1, 0) [event] {};
      \node (bg6) at (5.9, 0) [event] {};
      \node (bg4) at (8.4, 0) [event] {};
      \node (bg5) at (9.5, 0) [event] {};
    \end{scope}

    \begin{scope}[shift={(0, -2)}]
      \draw[->] (0, 0) node[left] {Gen.~1} -- (11, 0);
      \foreach \x in {0, 2, 4, 6, 8, 10}
      \draw[shift={(\x, 0)},color=black] (0pt, 3pt) -- (0pt, -3pt);

      \node (o1) at (1.8, 0) [triggered1] {};
      \node (o2) at (3.4, 0) [triggered1] {};
      \node (o3) at (4.3, 0) [triggered1] {};
      \node (o4) at (5.4, 0) [triggered1] {};

      \draw[->] (bg1) -- (o1);
      \draw[->] (bg1) -- (o2);
      \draw[->] (bg2) -- (o3);
      \draw[->] (bg3) -- (o4);
    \end{scope}

    \begin{scope}[shift={(0, -3)}]
      \draw[->] (0, 0) node[left] {Gen.~2} -- (11, 0);
      \foreach \x in {0, 2, 4, 6, 8, 10}
      \draw[shift={(\x, 0)},color=black] (0pt, 3pt) -- (0pt, -3pt);

      \node (o5) at (6.7, 0) [triggered2] {};
      \node (o6) at (2.6, 0) [triggered2] {};

      \draw[->] (o4) -- (o5);
      \draw[->] (o1) -- (o6);
    \end{scope}

    \begin{scope}[shift={(0, -4.5)}]
      \draw[->] (0, 0) node[left] {Labeled} -- (11, 0) node[right] {$t$};
      \foreach \x in {0, 2, 4, 6, 8, 10}
      \draw[shift={(\x, 0)},color=black] (0pt, 3pt) -- (0pt, -3pt) node[below] {$\x$};

      \foreach \x in {1.2, 3.2, 5.1, 5.9, 8.4, 9.5}
      \draw[shift={(\x, 0)}] (0,0) node[event] {};

      \foreach \x in {1.8, 3.4, 4.3, 5.4}
      \draw[shift={(\x, 0)}] (0,0) node[triggered1] {};

      \foreach \x in {2.6, 6.7}
      \draw[shift={(\x, 0)}] (0,0) node[triggered2] {};
    \end{scope}
  \end{tikzpicture}
  \caption{At top, a hypothetical observed self-exciting point process of events
    from \(t = 0\) to \(t = 10\). Below, the separation of that process into a
    background process and two generations of offspring processes. The arrows
    indicate the cluster relationships of which events were triggered by which
    preceding events; solid circles are background events, and open circles and
    squares are triggered events. At bottom, the combined process with
    generation indicated by shapes and shading. This cluster structure is not
    directly observed, though it may be inferred with the methods of
    Section~\ref{declustering}.}
  \label{branching-diagram}
\end{figure*}
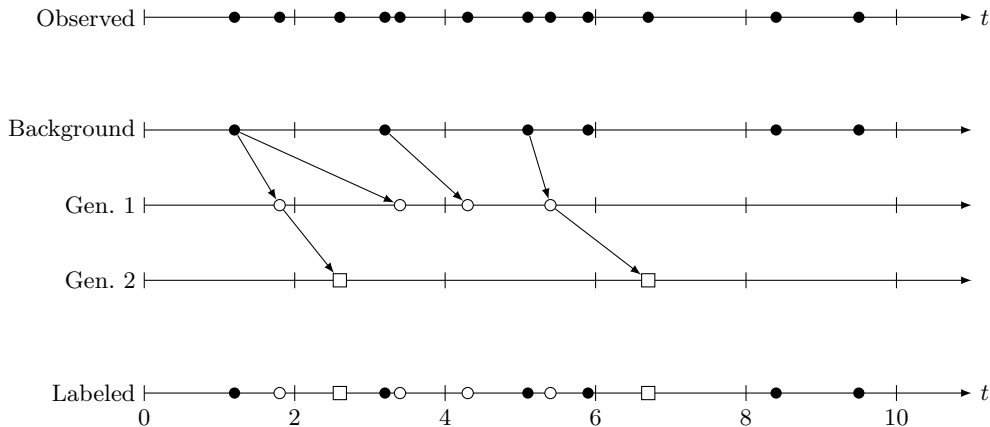

Hawkes processes have been put to many uses in a range of fields, modeling
financial transactions \citep{Bauwens:2009fk,Bacry_2015}, neuron activity
\citep{Johnson_1996}, terrorist attacks \citep{Porter:2012}, and a wide range of
other processes. They are particularly useful in processes that exhibit
clustering: \citet{Hawkes:1974ib} demonstrated that any stationary self-exciting
point process with finite intensity may be interpreted as a Poisson cluster
process. The events may be partitioned into disjoint processes: a
\textit{background process} of cluster centers \(N_c(t)\), which is simply a
Poisson process with rate \(\nu\), and separate \textit{offspring processes} of
triggered events inside each cluster, whose intensities are determined by \(g\).
Each triggered event may then trigger further events.
Fig.~\ref{branching-diagram} illustrates this separation. The number of
offspring of each event is drawn from a Poisson distribution with mean
\[
  m = \int_0^\infty g(t) \dif t.
\]
Provided \(m < 1\), cluster sizes are almost surely finite, as each generation
of offspring follows a geometric progression, with expected total cluster size
of \(1 / (1 - m)\) including the initial background event. This partitioning
also permits other useful results, such as an integral equation for the
distribution of the length of time between the first and last events of a
cluster \citep[Theorem 5]{Hawkes:1974ib}.

\subsection{Spatio-Temporal Form}
\label{spatiotemporal}

Spatio-temporal models extend the conditional intensity function to predict the
rate of events at locations \(s \in X \subseteq \mathbb{R}^d\) and times \(t \in
[0, T)\). The function is defined in the analogous way to temporal Hawkes
processes:
\begin{equation}\label{intensity-expectation}
  \lambda(s, t \mid {\cal H}_t) = \lim_{\Delta s, \Delta t \to 0} \frac{\E\left[ N\left(B(s,
        \Delta s)  \times [t, t + \Delta t)\right) \mid {\cal
        H}_t\right]}{|B(s, \Delta s)| \Delta t},
\end{equation}
where \(N(A)\) is again the counting measure of events over the set \(A
\subseteq X \times [0,T)\) and \(|B(s, \Delta s)|\) is the Lebesgue measure of
the ball \(B(s, \Delta s)\) with radius \(\Delta s\).

A \textit{self-exciting} spatio-temporal point process is one whose conditional
intensity is of the form
\begin{equation}\label{self-exciting}
  \lambda(s, t \mid {\cal H}_t) = \mu(s) + \sum_{i: t_i < t} g(s - s_i, t - t_i),
\end{equation}
where \(\{s_1, s_2, \dots, s_n\}\) denotes the observed sequence of locations of
events and \(\{t_1, t_2, \dots, s_n\}\) the observed times of these events.
Generally the triggering function \(g\) is nonnegative, and is often a kernel
function or power law decay function; often, for simplicity, it is taken to be
separable in space and time, so that \(g(s - s_i, t - t_i) = f(s - s_i) h(t -
t_i)\), similar to covariance functions in other spatio-temporal models
\citep[Section~6.1]{Cressie:2011vw}. Sometimes a general nonparametric form is
used, as in the model described in Section~\ref{misd}.

For ease of notation, the explicit conditioning on the past history \({\cal
  H}_t\) will be omitted for the rest of this review, and should be read as
implied for all self-exciting conditional intensities.

As with Hawkes processes, spatio-temporal self-exciting processes can be treated
as Poisson cluster processes, with the mean number of offspring
\begin{equation}\label{mean-offspring}
  m = \int_X \int_0^T g(s, t) \dif t \dif s.
\end{equation}
The triggering function \(g\), centered at the triggering event, is the
intensity function for the offspring process. Properly normalized, it induces a
probability distribution for the location and times of the offspring events. The
cluster process representation will prove crucial to the efficient estimation
and simulation of self-exciting processes, and the estimation of the cluster
structure of the process will be the focus of Section~\ref{declustering}.

\begin{figure*}
  \centering
  \includegraphics[width=0.9\textwidth]{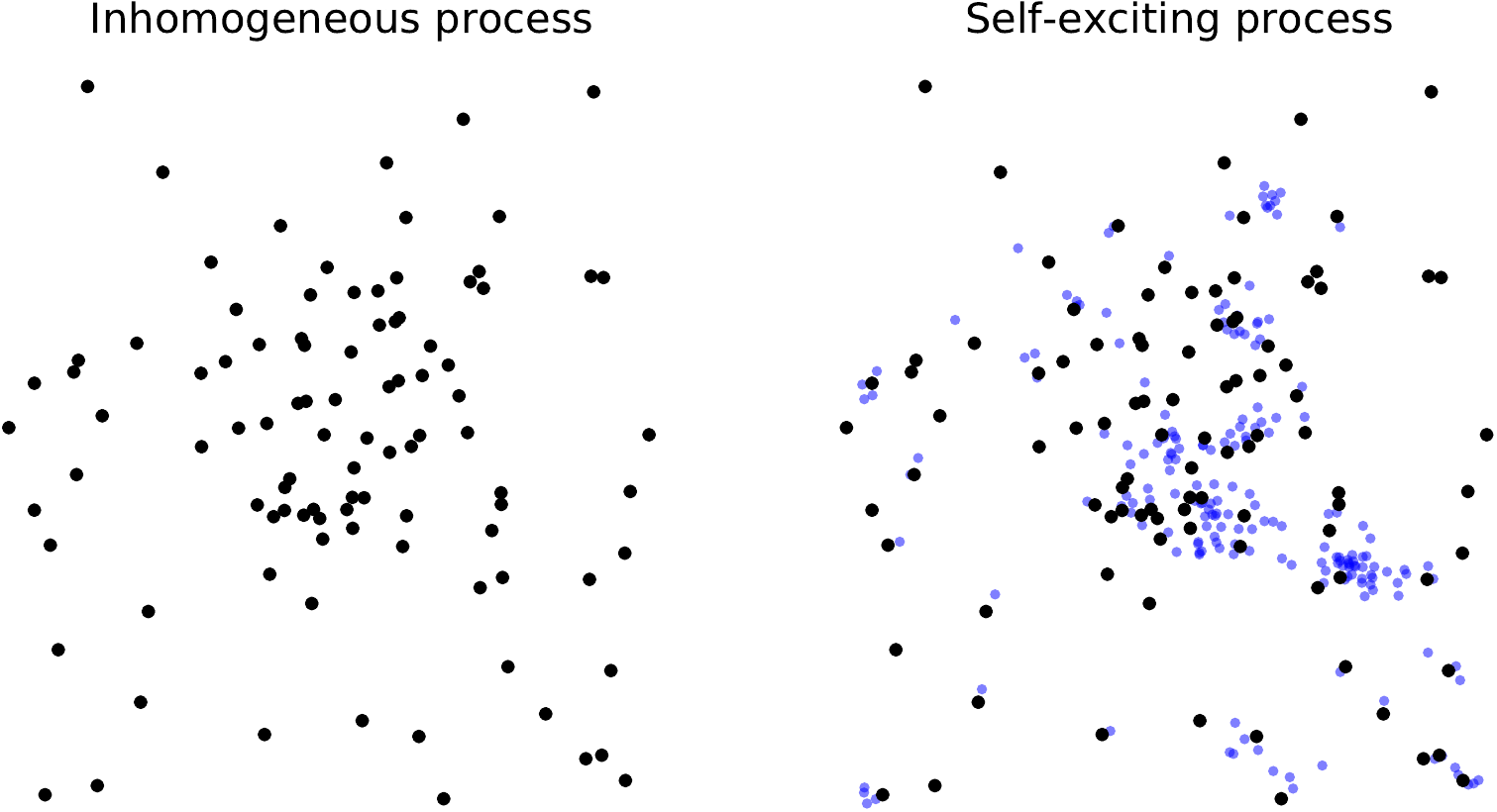}
  \caption{At left, a realization of an inhomogeneous Poisson process, in which
    the intensity is higher inside a central square and lower outside. At right,
    a self-exciting process with average total cluster size of 4, using the
    inhomogeneous Poisson process as the background process. Excited events are
    shown in blue. The cluster structure of the process is clearly visible, with
    clumps emerging from the self-excitation.}
  \label{compare-processes}
\end{figure*}

To illustrate the cluster process behavior of spatio-temporal self-exciting
processes, Fig.~\ref{compare-processes} compares a simulated realization of a
spatio-temporal inhomogeneous Poisson process against a self-exciting process
using the same Poisson process realization as its background process. The
self-exciting process, simulated using a Gaussian triggering function with a
short bandwidth, shows clusters (of expected total cluster size 4) emerging from
the Poisson process. The simulation was performed using
Algorithm~\ref{cluster-simulation}, to be discussed in Section~\ref{simulation},
which directly uses the cluster process representation to make simulation more
efficient.

\subsection{Marks}

Point processes may be \textit{marked} if features of events beyond their time
or location are also observed \citep[Section~6.4]{Daley:2003v1}. For example, if
earthquakes are treated as a spatiotemporal point process of epicenter locations
and times, the magnitude of each earthquake is an additional observed variable
which is an important part of the process: the number and distribution of
aftershocks may depend upon it. A marked point process is hence a point process
of events \(\{(s_i, t_i, \kappa_i)\}\), where \(s_i \in X \subseteq
\mathbb{R}^d\), \(t_i \in [0, T)\), and \(\kappa_i \in {\cal K}\), where \({\cal
  K}\) is the \textit{mark space} (e.g.\ the space of earthquake magnitudes). A
special case is the \textit{multivariate point process}, in which the mark space
is a finite set \(\{1, \dots, m\}\) for a finite integer \(m\). Often the mark
in a multivariate point process indicates the type of each event, such as the
type of crime reported.

Marks can have several useful properties. A process has \textit{independent
  marks} if, given the locations and times \(\{(s_i, t_i)\}\) of events, the marks
are mutually independent of each other, and the distribution of \(\kappa_i\)
depends only on \((s_i, t_i)\). Separately, a process has \textit{unpredictable
  marks} if \(\kappa_i\) is independent of all locations and marks \(\{(s_j,
t_j, \kappa_j)\}\) of previous events (\(t_j < t_i\)).

A marked point process has a \textit{ground process}, the point process of event
locations and times without their corresponding marks. Using the ground process
conditional intensity \(\lambda_g(s, t)\), we can write the marked point
process's conditional intensity function as
\begin{equation}\label{marked-intensity}
  \lambda(s, t, \kappa) = \lambda_g(s, t) f(\kappa \mid s, t),
\end{equation}
where \(f(\kappa \mid s, t)\) is the conditional density of the mark at time
\(t\) and location \(s\) given the history of the process up to \(t\). In
general, the ground process may depend on the past history of marks as well as
the past history of event locations and times. For simplicity of notation, the
following sections will largely consider point processes without marks, except
where noted, but most methods apply to marked and unmarked processes alike.

\subsection{Log-Likelihood}

The likelihood function for a particular parametric conditional intensity model
is not immediately obvious: given the potentially complex dependence caused by
self-excitation, even the distribution of the total number of events observed in
a time interval is difficult to obtain, and the spatial distributions of this
varying number of events must also be accounted for. Instead, for a realization
of \(n\) points from a point process, we start with its Janossy density
\citep[Section~5.3]{Daley:2003v1}. For a temporal point process, where a
realization is the set of event times \(\{t_1, t_2, \dots, t_n\}\) in a set
\(T\), the Janossy density is defined by the Janossy measure \(J_n\),
\[
  J_n(A_1 \times \cdots \times A_n) = n! p_n \Pi_n^\text{sym} (A_1 \times \cdots
  \times A_n),
\]
where the total number of events is \(n\), \(p_n\) is the probability of a
realization of the process containing exactly \(n\) events, \((A_1, \dots,
A_n)\) is a partition of \(T\) where \(A_i\) represents possible times for event
\(i\), and \(\Pi_n^\text{sym}(\cdot)\) is a symmetric probability measure
determining the joint distribution of the times of events in the process, given
there are \(n\) total events. The Janossy measure is not a probability measure:
it represents the sum of the probabilities of all \(n!\) permutations of \(n\)
points. It is nonetheless useful, as its density \(j_n(t_1, \dots, t_n) \dif t_1
\cdots \dif t_n\) has an intuitive interpretation as the probability that there
are exactly \(n\) events in the process, one in each of the \(n\) infinitesimal
intervals \((t_i, t_i + \dif t_i)\).

This interpretation connects the Janossy density to the likelihood function,
which can be written as \citep[Definition~7.1.II]{Daley:2003v1}
\begin{equation}\label{likelihood-janossy}
  L_T(t_1, \dots, t_n) = j_n(t_1, \dots, t_n \mid T)
\end{equation}
for a process on a bounded Borel set of times \(T\); for simplicity in the rest
of this section, we'll consider times in the interval \([0,T)\). Here \(j_n(t_1,
\dots, t_n \mid T)\) denotes the \emph{local} Janossy density, interpreted as
the probability that there are exactly \(n\) events in the process before time
\(T\), one in each of the infinitesimal intervals.

The likelihood can be rewritten in terms of the conditional intensity function,
which is usually easier to define than the Janossy density, by connection with
survival and hazard functions. Consider the conditional survivor functions
\(S_k(t \mid t_1, \dots, t_{k-1}) = \Pr(t_k > t \mid t_1, \dots, t_{k-1})\).
Using these functions and the conditional probability densities \(p_k(t \mid
t_1, \dots, t_{k-1})\) of event times, we can write the Janossy density
recursively as
\begin{equation}\label{janossy-recursive}
  \begin{split}
    j_n(t_1, \dots, t_n \mid T) ={}& p_1(t_1) p_2(t_2 \mid t_1) \cdots p_n(t_n
    \mid t_1, \dots, t_{n-1}) \times{}\\
    &S_{n+1}(T \mid t_1, \dots, t_n).
  \end{split}
\end{equation}
Additionally, we may define the hazard functions
\begin{equation}\label{hazardfn}
  \begin{split}
    h_k(t \mid t_1, \dots, t_{k_1}) &= \frac{p_k(t \mid t_1, \dots, t_{k-1})}{S_k(t
      \mid t_1, \dots, t_{k-1})}\\
    &= -\dod{\log S_k(t \mid t_1, \dots, t_{k-1})}{t}.
  \end{split}
\end{equation}
The hazard function has a natural interpretation as the conditional
instantaneous event rate---which means the conditional intensity \(\lambda(t)\)
can be written directly in terms of the hazard functions:
\[
  \lambda(t) = \begin{cases}
    h_1(t) & 0 < t < t_1\\
    h_k(t \mid t_1, \dots, t_{k-1}) & t_{k-1} < t \leq t_k, k \geq 2.
  \end{cases}
\]

This allows us to write the likelihood from eq.~\eqref{likelihood-janossy} in
terms of the conditional intensity function instead of the Janossy density.
Observe that from eq.~\eqref{hazardfn} we may write
\[
  S_k(t \mid t_1, \dots, t_{k-1}) = \exp\left(- \int_{t_{k-1}}^t h_k(u \mid t_1,
    \dots, t_{k-1}) \dif u\right)
\]
Substituting eq.~\eqref{hazardfn} into eq.~\eqref{janossy-recursive}, replacing
the hazard function with the conditional intensity, and combining terms leads to
the likelihood, for a complete parameter vector \(\Theta\), of
\citep[Proposition~7.2.III]{Daley:2003v1}
\[
  L(\Theta) = \left[ \prod_{i=1}^n \lambda(t_i) \right] \exp\left( - \int_0^T
    \lambda(t) \dif t\right).
\]
By treating spatial locations as marks, we may obtain extend this argument to
spatio-temporal processes and obtain the log-likelihood
\citep[Proposition~7.3.III]{Daley:2003v1}:
\begin{equation}\label{loglik}
  \ell(\Theta) = \sum_{i=1}^n \log\left(\lambda(s_i, t_i)\right) - \int_0^T \int_X
  \lambda(s, t) \dif s \dif t,
\end{equation}
where \(X\) is the spatial domain of the observations. For spatio-temporal
marked point processes with intensity defined as in
eq.~\eqref{marked-intensity}, the log-likelihood is written in terms of the
ground process, and has an extra mark term
\citep[Proposition~7.3.III]{Daley:2003v1}:
\[
  \begin{split}
    \ell(\Theta) ={}& \sum_{i=1}^n \log\left(\lambda_g(s_i, t_i)\right) + \sum_{i=1}^n \log \left(
      f(m_i \mid s_i, t_i) \right)\\
    &{}- \int_0^T \int_X \lambda_g(s, t) \dif s \dif t.
  \end{split}
\]

In unmarked processes, the first term in eq.~\eqref{loglik} is easy to
calculate, assuming the conditional intensity is straightforward, but the second
term can require computationally expensive numerical integration methods.

There are several approaches to evaluate this integral. The spatial domain \(X\)
can be arbitrary---e.g. a polygon defining the boundaries of a city---so
\citet{Meyer:2011ct} (see Section~\ref{epidemic-forecasting}) used
two-dimensional numeric integration via cubature, as part of a numerical
maximization routine. This requires an expensive numeric integration at every
step of the numerical maximization, making the procedure unwieldy.

\citet{Schoenberg:2013ab} observed that, for some conditional intensities, it may
be much easier to analytically integrate over \(\mathbb{R}^2\) instead of an
arbitrary \(X\). Hence the approximation
\[
  \int_0^T \int_X \lambda(s, t) \dif s \dif t \leq \int_0^\infty
  \int_{\mathbb{R}^2} \lambda(s, t) \dif s \dif t
\]
may reduce the integral to a form which may be evaluated directly. The
approximation is exact when the effect of self-excitation is contained entirely
within \(X\) and before \(t = T\), and overestimates otherwise; because
overestimation decreases the calculated log-likelihood, Schoenberg argued that
likelihood maximization will avoid parameter values where overestimation is
large. \citet{Lippiello_2014} argued that the temporal approximation biases
parameter estimates more than the spatial one, and advocated only approximating
\(X\) by \(\mathbb{R}^2\). This approximation was used by \citet{Mohler:2014iu},
discussed in Section~\ref{crime-forecasting}. \citet{Lippiello_2014} also
proposed a more accurate spatial approximation method based on a transformation
of the triggering function to polar coordinates.

\section{Estimation and Inference}
\label{estimation-inference}

Suppose now we have observed a realization of a self-exciting point process,
with event locations \(\{s_1, s_2, \dots, s_n\}\) and times \(\{t_1, t_2, \dots,
t_n\}\) over a spatial region \(X\) and temporal window \([0, T)\). We have a
model for the conditional intensity function and would like to be able to
estimate its parameters, perform inference, and simulate new data if needed.
This section discusses common approaches to these problems in the literature,
focusing largely on maximum likelihood estimation, though with a brief
discussion of Bayesian approaches in Section~\ref{bayesian}.

Fitting conditional intensity functions is not the only way to approach
spatio-temporal point processes; there is also extensive literature which
primarily uses descriptive statistics, such as first and second order moments of
the process. I will not delve into this literature here, as it is less useful
for understanding self-exciting processes; nonetheless, \citet{VereJones:2008iy}
gives a brief review, and more thorough treatments are available from
\citet{Gonzalez:2016fl} and \citet{Diggle:2014}.

\subsection{Maximum Likelihood}
\label{maximum-likelihood}

Self-exciting point process models are most commonly fit using maximum
likelihood. This is usually impossible to perform analytically: the form of the
log-likelihood in eq.~\eqref{loglik} involves a sum of logarithms of conditional
intensities, which themselves involve sums over previous points, making
analytical maximization intractable. Numerical evaluation of the intensity takes
\(O(n^2)\) time, and the log-likelihood can be nearly flat in large regions of
the parameter space, causing problems for numerical maximization algorithms and
making convergence extremely slow; in some examples explored by
\citet{Veen:2008eb}, numerical maximization may fail to converge altogether.
Nonetheless, for small datasets where the log-likelihood is computationally
tractable to evaluate, numerical maximization is often used.

Alternately, \citet{Veen:2008eb} showed the likelihood can be maximized with the
expectation maximization (EM) algorithm \citep{Dempster:1977,McLachlan:2008} by
introducing a latent quantity \(u_i\) for each event \(i\), which indicates
whether the event came from the background (\(u_i = 0\)) or was triggered by a
previous event \(j\) (\(u_i = j\)). This follows naturally from the cluster
process representation discussed in Sections~\ref{hawkes} and
\ref{spatiotemporal}: if \(u_i = 0\), event \(i\) is a cluster center, and
otherwise it is the offspring (directly or indirectly) of a cluster center.

\citet{Veen:2008eb} derived the complete-data log-likelihood for a specific
earthquake clustering model. More generally, consider a model of the form given
in eq.~\eqref{self-exciting}. If the branching structure \(u_i\) is assumed to
be known, the complete-data log-likelihood for a parameter vector \(\Theta\) can
be written as
\begin{equation*}
  \begin{split}
    \ell_c(\Theta) ={}& \sum_{i=1}^n \ind(u_i = 0)\log\left(\mu(s_i)\right)\\
      &{}+ \sum_{i=1}^n \sum_{j=1}^n \ind(u_i = j) \log\left(g(s_i - s_j, t_i -
        t_j)\right)\\
    &{}- \int_0^T \int_X \lambda(s, t) \dif s \dif t,
  \end{split}
\end{equation*}
where \(\ind(\cdot)\) is the indicator function, which is one when its argument
is true and zero otherwise. The branching structure dramatically simplifies the
log-likelihood, as each event's intensity comes only from its trigger (the
background or a previous event); this is analogous to the common EM approach to
mixture models, where the latent variables indicate the underlying distribution
from which each point came.

To complete the E step, we take the expectation of \(\ell_c(\Theta)\). This
requires estimating the triggering probabilities \(\Pr(u_i = j) = \E[\ind(u_i =
j)]\) for all \(i\), \(j\), based on the current parameter values \(\hat
\Theta\) for this iteration. We can calculate these probabilities as
\begin{align}\label{puij}
  \Pr(u_i = j) &= \begin{cases}
    \frac{g(s_i - s_j, t_i - t_j)}{\lambda(s_i, t_i)} & t_j < t_i\\
    0 & t_j \geq t_i
  \end{cases}\\\label{pui0}
  \Pr(u_i = 0) &= 1 - \sum_{j = 1}^{i - 1} P(u_i = j) =
               \frac{\mu(s_i)}{\lambda(s_i, t_i)}.
\end{align}
This leads to the expected complete-data log-likelihood
\begin{equation*}
  \begin{split}
    \E[\ell_c(\Theta)] ={}& \sum_{i=1}^n \Pr(u_i = 0) \log\left(\mu(s_i)\right)\\
    &{}+ \sum_{i=1}^n \sum_{j=1}^n \Pr(u_i = j) \log\left(g(s_i - s_j, t_i - t_j)\right)\\
    &{} - \int_0^T \int_X \lambda(s, t) \dif s \dif t,
  \end{split}
\end{equation*}
which is much easier to analytically or numerically maximize with respect to
each parameter in the M step. Once new parameter estimates are found, the
procedure returns to the E step, estimating new triggering probabilities, and
repeats until the log-likelihood converges, or until the estimated parameter
values change by less than some pre-specified tolerance.

The EM algorithm has several advantages over other numerical maximization
methods. Introducing the branching structure avoids the typical numerical issues
encountered by other maximization algorithms, making the maximization at each
iteration much easier, and the triggering probabilities also have a dual use in
stochastic declustering algorithms, discussed in the next section.

One important warning must be kept in mind, however. If we have observed only
data in the region \(X\) and time interval \([0, T)\), but the underlying
process extends outside this region and time, our parameter estimates will be
biased by boundary effects \citep{Zhuang:2004ex}. Unobserved events just outside
\(X\) or before \(t = 0\) can produce observed offspring which may be
incorrectly attributed to the background process, and observed events near the
boundary can produce offspring outside it, leading estimates of the mean number
of offspring \(m\) (see eq.~\eqref{mean-offspring}) to be biased downward.
Boundary effects can also bias the estimated intensity \(\lambda(s,t)\) in ways
analogous to the bias experienced in kernel density estimation
\citep{Cowling:1996kd}, but these effects are not well characterized for common
self-exciting models.

\subsection{Stochastic Declustering}
\label{declustering}

For some types of self-exciting point processes, the background event rate
\(\mu(s)\) is fit nonparametrically from the observed data, for example by
kernel density estimation or using splines \citep{Ogata:1988}. This could be fit
by maximum likelihood---\citet{Mohler:2014iu} fit the background as a weighted
kernel density via maximum likelihood, for example---but in some cases, we would
like to estimate \(\mu(s)\) using events from the background process only, and
not using events which were triggered by those events. We may also want to
analyze the background process intensity separately from the triggered events,
since the background process may have an important physical interpretation. This
requires a procedure which can separate background events from triggered events,
as illustrated in Fig.~\ref{branching-diagram}: stochastic declustering.

\subsubsection{Model-Based Stochastic Declustering.}

This version of stochastic declustering, introduced by \citet{Zhuang:2002fp},
assumes that the triggering function \(g\) has a parametric form, but that the
background \(\mu(s)\) should be estimated nonparametrically from only background
events. Estimating the background requires determining whether each event was
triggered by the background, but to do so requires \(g\), so the procedure is
iterative, starting with initial parameter values and alternately updating the
background estimate and \(g\) until convergence.

Consider the total spatial intensity function, defined as \citep{Zhuang:2002fp}
\begin{equation}\label{total-intensity}
  m_1(s) = \lim_{T \to \infty} \frac{1}{T} \int_0^T \lambda(s, t) \dif t,
\end{equation}
where \(T\) is the length of the observation period. The function \(m_1(s)\)
does not require declustering to estimate, since it sums over all events,
including triggered events; by replacing the limit in
eq.~\eqref{total-intensity} with a finite-data approximation and substituting in
eq.~\eqref{self-exciting}, we obtain
\begin{align*}
  m_1(s) &\approx \frac{1}{T} \int_0^T \mu(s) + \sum_{i:t_i < t} g(s - s_i, t - t_i)
           \dif t\\
         &= \mu(s) + \frac{1}{T} \int_0^T \sum_{i:t_i < t} g(s - s_i, t - t_i)
           \dif t.
\end{align*}
We hence obtain the relation
\begin{equation}\label{mu-approx}
  \mu(s) \approx m_1(s) - \frac{1}{T} \sum_{i:t_i < t} \int_0^T g(s - s_i, t -
  t_i) \dif t.
\end{equation}

We can now use a suitable nonparametric technique, such as kernel density
estimation, to form \(\hat m_1(s)\):
\[
  \hat m_1(s) = \frac{1}{T} \sum_{i=1}^n k(s - s_i),
\]
where \(k\) is a kernel function. It may also be desirable to estimate the
second term on the right-hand side of eq.~\eqref{mu-approx}, denoted
\(\gamma(s)\), the same way. To do so, we use the same latent quantity \(u_i\)
defined and estimated in Section~\ref{maximum-likelihood}. We can estimate the
cluster process by, for example, a weighted kernel density estimate, using
\[
  \hat \gamma(s) = \frac{1}{T} \sum_{i=1}^n \Pr(u_i \neq 0) k(s - s_i).
\]
This leads to the estimator
\begin{equation}\label{mu-estimator}
  \begin{split}
    \hat \mu(s) &= \hat m_1(s) - \hat \gamma(s)\\
    &= \frac{1}{T} \sum_{i=1}^n (1 - \Pr(u_i \neq 0)) k(s - s_i).
  \end{split}
\end{equation}

We now need to iteratively estimate parameters of the triggering function \(g\).
Provided these can be found by maximum likelihood, \citet{Zhuang:2002fp}
suggested the following algorithm:
\begin{algorithm}\label{iterative-fit}
  Let \(\hat \mu(s) = 1\) initially.
  \begin{enumerate}
  \item Using maximum likelihood (see Section~\ref{maximum-likelihood}), fit the
    parameters of the conditional intensity function
    \[
      \lambda(s, t) = \hat \mu(s) + \sum_{i:t_i < t} g(s - s_i, t - t_i).
    \]
  \item Calculate \(\Pr(u_i \neq 0)\) for all \(i\) using the parameters found in
    step 1 and eq.~\eqref{pui0}.
  \item Using the new branching probabilities, form a new \(\hat \mu^*(s)\)
    using eq.~\eqref{mu-estimator}.
  \item If \(\max_s |\hat \mu(s) - \hat \mu^*(s)| > \epsilon\), for a pre-chosen
    tolerance \(\epsilon > 0\), return to step 1. Otherwise, terminate the
    algorithm.
  \end{enumerate}
\end{algorithm}

We can now perform stochastic declustering by thinning the process. With the
final estimated \(\hat \mu(s)\), we recalculate \(\Pr(u_i \neq 0)\) and keep each
event with probability \(1 - \Pr(u_i \neq 0)\); the rest of the events are
considered triggered events and deleted. We are left with those identified as
background events.

In the original implementation of this algorithm, \citet{Zhuang:2002fp} used an
adaptive kernel function \(k\) in eq.~\eqref{mu-estimator} whose bandwidth was
chosen separately for each event, rather than being uniform for the whole
dataset. After choosing an integer \(n_p\) between 10 and 100, for each event
they found the smallest disk centered at that event which includes at least
\(n_p\) other events (forced to be larger than some small value \(\epsilon\),
chosen on the order of the observation error in locations). The radius of this
disk was used as the bandwidth for the kernel at each event. This method was
chosen because, in clustered datasets, any single bandwidth oversmooths in some
areas and is too noisy in others. A method to estimate kernel parameters from
the data will be introduced in Section~\ref{flp}.

\citet{Zhuang:2002fp} also adapted the declustering algorithm to produce a
``family tree'': a tree connecting background events to the events they trigger,
and so on from each event to those it triggered. The algorithm considers each
pair of events and determines whether one should be considered the ancestor of
the other:
\begin{algorithm}\label{declustering-tree}
  Begin with the final estimated \(\hat \mu(s)\) from
  Algorithm~\ref{iterative-fit}.
  \begin{enumerate}
  \item For each pair of events \(i\), \(j\) (with \(t_i > t_j\)), calculate
    \(\Pr(u_i = j)\) and \(\Pr(u_i = 0)\).
  \item Set \(i = 1\).
  \item Generate a uniform random variate \(R_i \sim \text{Uniform}(0, 1)\).
  \item If \(R_i < \Pr(u_i = 0)\), consider event \(i\) to be a background event.
  \item Otherwise, select the smallest \(J\) such that \(R_i < \Pr(u_i = 0) +
    \sum_{j=1}^J \Pr(u_i = j)\). Consider the \(i\)th event to be a descendant of
    the \(J\)th event.
  \item When \(i = N\), the total number of events, terminate; otherwise, set
    \(i = i + 1\) and return to step 3.
  \end{enumerate}
\end{algorithm}

Though the thinning algorithm and family tree construction are stochastic and
hence do not produce unique declusterings, \citet{Zhuang:2002fp} argue this is
an advantage, as uncertainty in declustering can be revealed by running the
declustering process repeatedly and examining whether features are consistent
across declustered processes. These methods have been used to answer important
scientific questions in seismology, discussed in Section~\ref{etas}.

\subsubsection{Forward Likelihood-based Predictive approach.}
\label{flp}

In a semiparametric model, where the background \(\mu(s)\) is estimated
nonparametrically from background events, the nonparametric estimator (such as a
kernel smoother) may have tuning parameters which need to be adapted to the
data. The model-based stochastic declustering procedure discussed above uses an
adaptive kernel in \(\mu(s)\), but we may wish to use a standard kernel density
estimator with bandwidth estimated from the data. However, if we follow
Algorithm~\ref{iterative-fit}, adjusting the bandwidth with maximum likelihood
at each iteration, the bandwidth would go to zero, placing a point mass at each
event.

To avoid this problem, \citet{Chiodi_2011} introduced the Forward
Likelihood-based Predictive approach (FLP). Rather than directly maximizing the
likelihood, consider increments in the log-likelihood, using the first \(k\)
observations to predict the \((k+1)\)th:
\begin{equation*}
  \begin{split}
    \delta_{k,k+1} (\Theta \mid {\cal H}_{t_k}) ={}& \log \lambda(s_{k+1},
    t_{k+1} \mid \Theta, {\cal H}_{t_k})\\
    &{}- \int_{t_k}^{t_{k+1}} \int_X \lambda(s, t \mid \Theta, {\cal H}_{t_k})
    \dif s \dif t,
  \end{split}
\end{equation*}
where the past history \({\cal H}_{t_k}\) explicitly indicates that the
intensity experienced by point \(k+1\) depends only on the first \(k\)
observations (i.e.\ the estimate of \(\mu(s)\) only includes the first \(k\)
points). A parameter estimate \(\hat \Theta\) is formed by numerically
maximizing the sum
\[
  \text{FLP}(\hat \Theta) = \sum_{k=k_1}^{n-1} \delta_{k,k+1}(\hat \Theta \mid
  {\cal H}_{t_k}),
\]
where \(k_1 = \lfloor n/2 \rfloor\). \citet{Adelfio_2014} and
\citet{Adelfio_2015} developed the FLP method into a semiparametric method
following an alternated estimation procedure similar to
Algorithm~\ref{iterative-fit}. The procedure splits the model parameters into
the nonparametric smoothing parameters \(\Sigma\) and the triggering function
parameters \(\Theta\), and iteratively fits them in the following steps:
\begin{algorithm}
  Begin with a default estimate for \(\Sigma\), for example by Silverman's rule
  for kernel bandwidths \citep{Silverman:1986up}. Use this to estimate
  \(\mu(s_i)\) for each event \(i\).
  \begin{enumerate}
  \item Using the estimated values of \(\mu(s_i)\) and holding \(\Sigma\) fixed,
    estimate the triggering function parameters \(\Theta\) via maximum
    likelihood.
  \item Calculate \(\Pr(u_i = 0)\) for each event \(i\) using the current
    parameter estimates.
  \item Estimate the smoothing parameters by maximizing \(\text{FLP}(\hat
      \Sigma)\), holding \(\Theta\) fixed.
  \item Calculate new estimates of \(\mu(s_i)\) for each event \(i\), using a
    weighted estimator with the weights calculated in step 2.
  \item Check for convergence in the estimates of \(\Sigma\) and \(\Theta\) and
    either terminate or return to step 1.
  \end{enumerate}
\end{algorithm}

\citet{Adelfio_2015} applied this method to a large catalog of earthquakes in
Italy, using the earthquake models to be discussed in Section~\ref{etas},
finding improved performance over a version of the model where smoothing
parameters were fixed solely with Silverman's rule.

\subsubsection{Model-Independent Stochastic Declustering.}
\label{misd}

\citet{Marsan:2008ci} proposed a model-independent declustering algorithm (MISD)
for earthquakes which removed the need for a parametric triggering function
\(g(s, t)\), instead estimating the shape of \(g(s, t)\) from the data. They
assumed a conventional conditional intensity with constant background rate
\(\lambda_0\),
\[
  \lambda(s, t) = \lambda_0 + \sum_{i : t_i < t} g(s - s_i, t - t_i),
\]
but \(g(s, t)\) was simply assumed to be piecewise constant in space and time,
with the constant for each spatial and temporal interval estimated from the
data. \citet{Marsan:2010fq} showed their method can be considered an EM
algorithm, following the same steps as in Section~\ref{maximum-likelihood}:
estimate the probabilities \(\Pr(u_i = j)\) in the E step and then maximize over
parameters of \(g(s, t)\) and \(\lambda_0\) in the M step, eventually leading to
convergence and final estimates of the branching probabilities.

\citet{Fox:2016bl} extended this method to the case where the background
\(\lambda_0\) is not constant in space by assuming a piecewise constant
background function \(\mu(s)\) or by using a kernel density estimate of the
background, then quantified uncertainty in the background and in \(g(s, t)\) by
using a version of the parametric bootstrap method to be discussed in
Section~\ref{asymptotic-normality}. This can be considered a general
nonparametric approach to spatio-temporal point process modeling as well as a
declustering method, since with confidence intervals for the nonparametric
triggering function, useful inference can be drawn for the estimated triggering
function's shape.

\subsection{Simulation}
\label{simulation}

It is often useful to simulate data from a chosen model. For temporal point
processes, a range of simulation methods are described by \citet[section
7.5]{Daley:2003v1}. Several spatio-temporal methods are based on a thinning
procedure which first generates a large quantity of events, then thins them
according to their conditional intensity, starting at the first event and
working onward so history dependence can be taken into account. The basic method
was introduced for nonhomogeneous Poisson processes by \citet{Lewis:1979ij}.

\citet{Ogata:1998cx} proposed a two-stage algorithm for general self-exciting
processes which requires thinning fewer events and is hence more efficient.
Events are generated sequentially, and the time of each event is determined
before its location. To generate times, we require a version of the conditional
intensity which is only a function of time, having integrated out space:
\begin{align*}
  \lambda_X(t) &= \nu_0 + \sum_{j: t_j < t} \nu_j(t)\\
  \nu_0 &= \int_X \mu(s) \dif s\\
  \nu_j(t) &= \int_X g(s, t) \dif s.
\end{align*}
This allows us to simulate times of events before simulating their locations.
The algorithm below, though apparently convoluted, amounts to drawing the
waiting time until the next event from an exponential distribution, drawing its
location according to the distribution induced by \(g\), and repeating,
rejecting (thinning) some proposed times proportional to their intensities
\(\lambda_X\):

\begin{algorithm}
  Start with \(a = b = c = 0\) and \(i = 1\).

  \begin{enumerate}
  \item Set \(s_a = 0\) and generate \(U_b \sim \text{Uniform}(0, 1)\). Let
    \(\Lambda_c = \nu_0\) and \(u_a = - \log(U_b) / \Lambda_c\).
  \item If \(u_a > T\), stop. Otherwise, let \(t_i = u_a\), let \(J = 0\), and
    skip to step 7.
  \item Let \(b = b + 1\) and \(a = a + 1\). Generate \(U_b \sim
    \text{Uniform}(0, 1)\) and let \(u_a = - \log(U_b) / \Lambda_c\).
  \item Let \(s_a = s_{a-1} + u_a\). If \(s_a > T\), stop; otherwise let \(b = b
    + 1\) and generate \(U_b \sim \text{Uniform}(0, 1)\).
  \item If \(U_b > \lambda_X(s_a) / \Lambda_c\), set \(c = c + 1\) and let
    \(\Lambda_c = \lambda_X(s_a)\), then go to step 3.
  \item Let \(t_i = s_a\), set \(b = b + 1\), generate \(U_b \sim
    \text{Uniform}(0, 1)\), and find the smallest \(J\) such that \(\sum_{j =
      0}^J \nu_j(t_i) > U_b \lambda_X(t_i)\).
  \item If \(J = 0\) then generate \(s \in X\) from the non-homogeneous Poisson
    intensity \(\mu(s)\) and go to step 10.
  \item Otherwise, set \(b = b + 1\), then set \(s_i\) by drawing from the
    normalized spatial distribution of \(g\) centered at \(s_J\).
  \item If \(s_i\) is not in \(X\), return to step 3.
  \item Otherwise, set \(i = i + 1\) and return to step 3.
  \end{enumerate}
\end{algorithm}

This can be computationally expensive. The intensity \(\lambda_X\) must be
evaluated at each candidate point, involving a large sum, and the thinning in
step 5 means multiple candidate times will often have to be generated. Another
method, developed for earthquake models, directly uses the cluster structure of
the self-exciting process, eliminating the need for thinning or repeated
evaluation of \(\lambda(s, t)\) \citep{Zhuang:2004ex}:

\begin{algorithm}\label{cluster-simulation}
  Begin with a fully specified conditional intensity \(\lambda(s, t)\).

  \begin{enumerate}
  \item Generate events from the background process using the intensity
    \(\mu(s)\), by using a simulation method for nonhomogeneous stationary
    Poisson processes \citep[e.g.][]{Lewis:1979ij}. Call this catalog of events
    \(G^{(0)}\).
  \item Let \(l = 0\).
  \item For each event \(i\) in \(G^{(l)}\), simulate its \(N^{(i)}\) offspring,
    where \(N^{(i)} \sim \text{Poisson}(m)\) (with \(m\) defined as in
    eq.~\eqref{mean-offspring}), and the offspring's location and time are
    generated from the triggering function \(g\), normalized as a probability
    density. Call these offspring \(O_i^{(l)}\).
  \item Let \(G^{(l + 1)} = \union_{i \in G^{(l)}} O_i^{(l)}\).
  \item If \(G^{(l)}\) is not empty, set \(l = l + 1\) and return to step 3.
    Otherwise, return \(\union_{j=0}^l G^{(j)}\) as the final set of simulated
    events.
  \end{enumerate}
\end{algorithm}

This algorithm has been widely used in the seismological literature for studies
of simulated earthquake catalogs. However, both methods suffer from the same
edge effects as discussed in Section~\ref{maximum-likelihood}: if the background
is simulated over a time interval \([0, T)\), the offspring of events occurring
just before \(t = 0\) are not accounted for. Similarly, if events occurred just
outside the spatial region \(X\), they can have offspring inside \(X\), which
will not be simulated. This can be avoided by simulating over a larger
space-time window and then only selecting simulated events inside \(X\) and
\([0, T)\). \citet{Moller:2005ke} developed a perfect simulation algorithm for
temporal Hawkes processes which avoids edge effects, but its extension to
spatio-temporal processes remains to be developed.

\subsection{Asymptotic Normality and Inference}
\label{asymptotic-normality}

\citet{Ogata:1978dt} demonstrated asymptotic normality of maximum likelihood
parameter estimates for temporal point processes, and showed the covariance
converges to the inverse of the expected Fisher information matrix, suggesting
an estimator based on the Hessian of the log-likelihood at the maximum
likelihood estimate. This estimator has been frequently used for spatio-temporal
models in seismology; however, \citet{Wang:2010dw}, comparing it with sampling
distributions found by repeated simulation, found that standard errors based on
the Hessian can be heavily biased for small to moderate observation period
lengths, suggesting the finite-sample behavior is poor.

\citet{Rathbun:1996bf} later demonstrated that for spatio-temporal point
processes, maximum likelihood estimates of model parameters are consistent and
asymptotically normal as the observation time \(T \to \infty\), under regularity
conditions on the form of the conditional intensity function \(\lambda(s, t)\).
An estimator for the asymptotic covariance of the estimated parameters is
\begin{equation}\label{sigmahat}
  \hat \Sigma = \left( \sum_{i = 1}^n \frac{\Delta(s_i, t_i)}{\lambda(s_i, t_i)}
  \right)^{-1},
\end{equation}
where \(\Delta(s_i, t_i)\) is a matrix-valued function whose entries are
\[
  \Delta_{ij}(s, t) = \frac{\dot \lambda_i(s, t) \dot \lambda_j(s,
    t)}{\lambda(s, t)}
\]
and \(\dot \lambda_i(s, t)\) denotes the partial derivative of \(\lambda(s, t)\)
with respect to the \(i\)th parameter. From \(\hat \Sigma\) we can derive Wald
tests of parameters of interest, and by inverting the tests we can obtain
confidence intervals for any parameter.

Rather than relying on asymptotic normality, another approach is the parametric
bootstrap, which has been used for temporal point process models in neuroscience
\citep{Sarma:2011ix}. The parametric bootstrap, though computationally
intensive, is conceptually simple:
\begin{algorithm}
  Using the parameter values \(\hat \Theta\) from a previously fitted model, and
  starting with \(i = 1\):
  \begin{enumerate}
  \item Using a simulation algorithm from Section~\ref{simulation}, simulate a
    new dataset in the same spatio-temporal region.
  \item Fit the same model to this new data, obtaining new parameter values
    \(\hat \Theta^{(i)}\).
  \item Repeat steps 1 and 2 with \(i = i + 1\), up to some pre-specified number
    of simulations \(B\) (e.g 1000).

    (Alternately, the algorithm can be adaptive, by checking the confidence
    intervals after every \(b\) steps and stopping when they seem to have
    converged.)
  \item Calculate bootstrap 95\% confidence intervals for each parameter by
    using the 2.5\% and 97.5\% quantiles of the estimated \(\hat \Theta^{(i)}\).
  \end{enumerate}
\end{algorithm}

This is straightforward to implement, relies on minimal assumptions, and is
asymptotically consistent in some circumstances. However, just as asymptotically
normal standard errors may be biased for finite sample sizes, the bootstrap has
no performance guarantees on small samples. \citet{Wang:2010dw} tested neither
the parametric bootstrap nor the estimator of \citet{Rathbun:1996bf} in their
simulations, so no direct comparison is possible here, and those intending to
use the bootstrap should test its performance in simulation.

It is sometimes desirable to estimate only a subset of the parameters in a
model, either because full estimation is intractable or because some covariates
are unknown. Dropping terms from the conditional intensity results in a
\emph{partial} likelihood, and parameter estimates obtained by maximizing the
partial likelihood may differ from those obtained from the complete likelihood.
\citet{Schoenberg:2016vy} explored the circumstances under which the parameter
estimates are not substantially different, finding that partial likelihood
estimates are identical under assumptions about the separability of the omitted
parameters, and are still consistent in more general additive models under
assumptions that the omitted parameters have relatively small effects on the
intensity. In either case, the maximum partial likelihood estimates still have
the asymptotic normality properties discussed above.

\subsection{Bayesian Approaches}
\label{bayesian}

\citet{Rasmussen:2011ns} introduced two methods for Bayesian estimation for
self-exciting temporal point processes: direct Markov Chain Monte Carlo (MCMC)
on the likelihood, using Metropolis updates within a Gibbs sampler, and a method
based on the cluster process structure of the process. \citet{Loeffler:2016}
recently adapted MCMC to fit a version of the self-exciting crime model
discussed in Section~\ref{crime-forecasting}, using the Stan modeling language
\citep{Stan} and Hamiltonian Monte Carlo to obtain samples from the posteriors
of the parameters. \citet{Ross:2016}, however, working with the seismological
models discussed in Section~\ref{etas}, argued that direct Monte Carlo methods
are impractical: a sampling method involving repeated rejection requires
evaluating the likelihood many times, an \(O(n^2)\) operation, and the strong
correlation of some parameters can make convergence difficult.

Instead, building on the cluster process method suggested by
\citet{Rasmussen:2011ns}, \citet{Ross:2016} proposed taking advantage of the
same latent variable formulation introduced for maximum likelihood in
Section~\ref{maximum-likelihood}. If the latent \(u_i\)s are known for all
\(i\), events in the process can be partitioned into \(N+1\) sets \(S_0, \dots,
S_N\), where
\[
  S_j = \{t_i \mid u_i = j\}, \qquad 0 \leq j < N.
\]
Events in each set \(S_j\) can be treated as coming from a single inhomogeneous
Poisson process, with intensity proportional to the triggering function \(g\)
(or to \(\mu(s)\), for \(S_0\)). This allows the log-likelihood to be
partitioned, reducing dependence between parameters and dramatically improving
sampling performance. The algorithm now involves sampling \(u_i\) (using the
probabilities defined in eqs.~\eqref{puij}--\eqref{pui0}), then using these
to sample the other parameters, in a procedure very similar to the expectation
maximization algorithm for these models.

\subsection{Model Selection and Diagnostics}
\label{model-selection}

In applications, model selection is usually performed using the Akaike
information criterion (AIC) or related criteria like the Bayesian information
criterion (BIC) and the Hannan--Quinn criterion; \citet{Chen:2017} compared the
performance of these methods in selecting the correct model in a range of
settings and sample sizes, finding AIC more effective in small samples and less
in larger samples. A variety of tests and residual plots are available for
evaluating the fit of spatio-temporal point process models. \citet{Bray:2013fb}
provide a comprehensive review focusing on earthquake models; I will give a
brief summary here.

First, we observe that any process characterized by its conditional intensity
\(\lambda(s, t)\) may be thinned to obtain a homogeneous Poisson process
\citep{Schoenberg:2003jv}, allowing examination of the fit of the spatial
component of the model. We define \(b = \inf_{s, t} \lambda(s, t)\), and for
each event \(i\) in the observed process, calculate the quantity
\[
  p_i = \frac{b}{\lambda(s_i, t_i)}
\]
Retain event \(i\) with probability \(p_i\). If this is done with an estimated
intensity \(\hat \lambda(s, t)\) from the chosen model, the thinned process (now
ignoring time) will be Poisson with rate \(b\), and can be examined for
homogeneity, for example with the \(K\)-function \citep{Ripley:1977la}, which
calculates the proportion of events per unit area which are within a given
distance. This will detect if the thinned process still has clustering not
accounted for by the model.

If \(b\) is small, the thinned process will contain very few events, making the
test uninformative. \citet{Clements:2012bt} propose to solve this problem with
``super-thinning'', which superimposes a simulated Poisson process. We choose a
rate \(k\) for the super-thinned process, such that \(b \leq k \leq \sup_{s, t}
\lambda(s, t)\), and thin with probabilities
\[
  p_i = \min\left\{\frac{k}{\lambda(s_i, t_i)}, 1\right\}.
\]
We add to the thinned process a simulated inhomogeneous Poisson process with
rate \(\max\{k - \lambda(s, t), 0\}\). The sum process is, if the estimated
model is correct, homogeneous with rate \(k\).

Graphical diagnostics are also available. For purely spatial point processes,
\citet{Baddeley_2005} developed a range of residual diagnostic tools to display
differences between the fitted model and the data, demonstrating further
properties of these residuals in \citet{Baddeley_2007} and
\citet{Baddeley_2011}. \citet{Zhuang_2006} showed these tools could be extended
directly to spatio-temporal point processes, producing residual maps which
display the difference between the predicted number of events and the actual
number, over grid cells or some other division of space. \citet{Bray:2014eo}
argued that a grid is a poor choice: if grid cells are small, the expected
number of events per cell is low and the distribution of residuals is skewed,
but if grid cells are large, over- and under-prediction within a single cell can
cancel out. Instead, they proposed using the Voronoi tesselation of space: for
each event location \(s_i\), the corresponding Voronoi cell consists of all
points which are closer to \(s_i\) than to any other event. This generates a set
of convex polygons. By integrating the conditional intensity over a reasonable
unit of time and over each Voronoi cell, we obtain a map of expected numbers of
events, which we can subtract from the true number in each cell (which is 1 by
definition). This produces a map which can be visually examined for defects in
prediction.

\begin{figure}
  \centering
  \includegraphics[width=0.7\columnwidth]{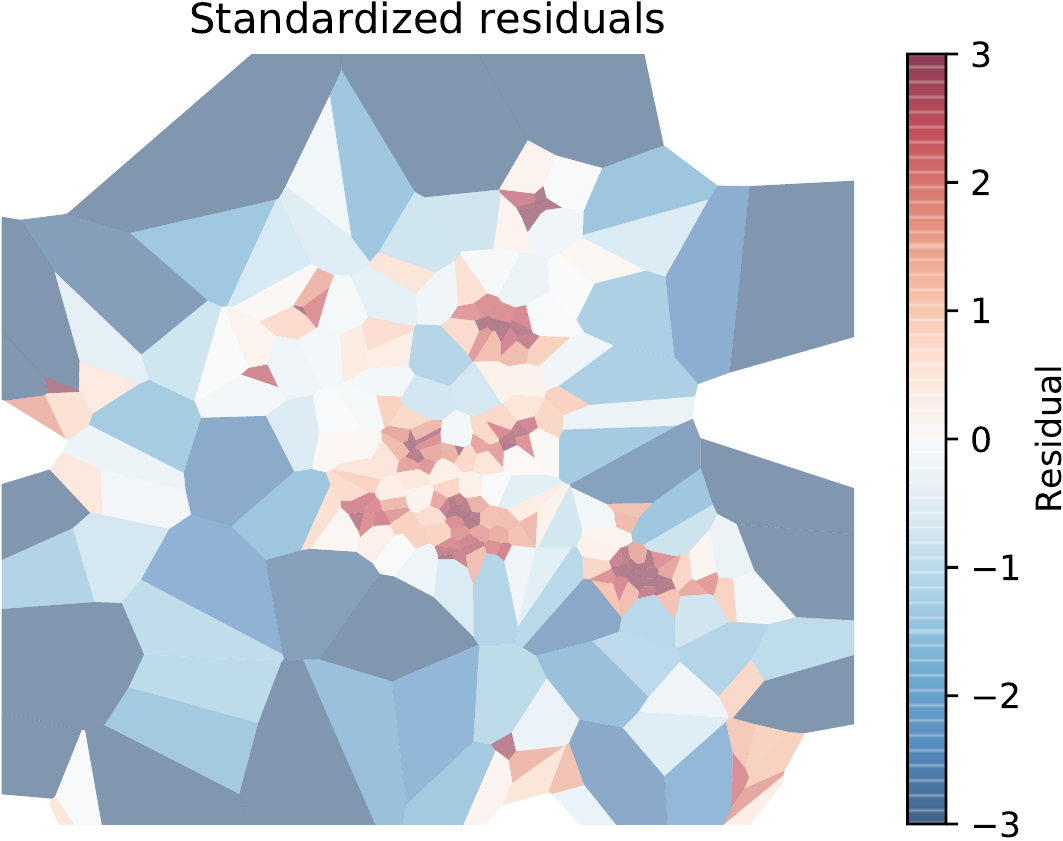}
  \caption{A Voronoi residual map of the self-exciting point process shown in
    Fig.~\ref{compare-processes}. The model was fit assuming a constant
    background intensity and does not account for the inhomogeneous rate,
    leading to positive residuals in the center area and negative residuals
    outside. Residual values are standardized according to an approximate
    distribution given by \citet{Bray:2014eo}.}
  \label{compare-resids}
\end{figure}

As an example, Fig.~\ref{compare-resids} is a Voronoi residual map of the
self-exciting point process previously shown in Fig.~\ref{compare-processes},
produced following the procedure suggested by \citet{Bray:2014eo}. A model was
fit to the simulated point process data which does not account for the
inhomogeneous background process, instead assuming a constant background rate,
and a spatial pattern in the residuals is apparent, with positive residuals
(more events than predicted) in areas where the background rate is higher and
negative residuals outside those areas.

\section{Applications}
\label{applications}

This section will review four major applications of self-exciting point
processes: earthquake models, crime forecasting, epidemic infection forecasting,
and events on networks. This is by no means an exhaustive list---self-exciting
point process models have been applied to problems as disparate as wildfire
occurrence \citep{Peng:2005bq} and civilian deaths in Iraq \citep{Lewis_2011}.
The selected applications illustrate the features that make self-exciting point
processes valuable: parameters of the triggering function \(g\) have important
physical interpretations and can be used to test scientific hypotheses about the
event triggering process, while the background \(\mu\) flexibly incorporates
spatial and temporal covariates whose effects can be estimated. Purely
descriptive methods, or methods such as log-Gaussian Cox processes, do not
permit the same inference about the event triggering process.

\subsection{Earthquake Aftershock Sequence Models}
\label{etas}

After a large earthquake, a sequence of smaller aftershocks is typically
observed in the days and weeks afterwards, usually near the epicenter of the
main shock \citep{Freed_2005}. These tremors are triggered by the seismic
disturbance of the main shock, and the distribution of their magnitudes and
arrival times has proven to be relatively consistent, allowing the development
of models for their prediction and analysis.

Sequences of earthquakes and aftershocks show rich behavior, such as spatial and
temporal clustering, complex spatial dependence, and gradual shifts in overall
seismicity. Self-exciting point processes are a natural choice to model this
behavior, as they can directly capture spatio-temporal aftershock triggering
behavior and can incorporate temporal trends and spatial inhomogeneity. The
Epidemic-Type Aftershock Sequence (ETAS) model, developed and expanded over
several decades, provides a flexible foundation for modeling this behavior, and
has been widely applied to earthquake sequences in Japan, California, and
elsewhere. A comprehensive review is provided by \cite{Ogata:1999jz}.

The initial ETAS model was purely temporal, modeling the rate of earthquakes at
time \(t\) as a superposition of a constant rate of background seismicity and of
aftershocks triggered by these background events:
\[
  \lambda(t) = \mu + \sum_{i:t_i < t} \frac{K_i}{(t - t_i + c)^p}
\]
Here \(\mu\) is the background seismic activity rate and \(K_i\) is related to
the recorded magnitude \(M_i\) of earthquake \(i\) by the relationship
\[
  K_i = K_0 e^{\alpha (M_i - M_0)},
\]
where \(M_0\) is the minimum magnitude threshold for earthquakes to be recorded
in the dataset, and \(K_0\), \(\alpha\), and \(p\) are constants. Earthquake
magnitudes are treated as unpredictable marks. The functional form of the
triggering function, known as the modified Omori formula, was determined
empirically by studies of aftershock sequences.

The temporal ETAS model was soon extended to a spatio-temporal model of the form
in eq.~\eqref{self-exciting}. A variety of triggering functions \(g\) were used,
ranging from bivariate normal kernels to more complicated exponential decay
functions and power laws; some triggering functions allow the range of spatial
influence to depend on the earthquake magnitude. The inhomogeneous background
\(\mu(s)\), which represents spatial differences in fault structure and tectonic
plate physics, can be obtained by a simple kernel density estimate
\citep{Musmeci_1992} or by the stochastic declustering methods discussed in
Section~\ref{declustering}.

\citet{Zhuang:2004ex} demonstrated that stochastic declustering can be used to
test model assumptions. They applied the ETAS model and stochastic declustering
to a catalog of 19,139 earthquakes compiled by the Japanese Meteorological
Agency, then used the declustered data to test assumptions typically used in
modeling earthquakes; for example, the distribution of earthquake magnitudes is
assumed to be the same for main shocks and aftershocks, and both mainshocks and
aftershocks trigger further aftershocks with the same spatial and temporal
distribution. By identifying main shocks and aftershocks and connecting them
with their offspring, it was possible to test each assumption, finding that some
do not hold and leading to a revised model \citep{Ogata:2006}.

Further, by using AIC, different triggering functions have been compared to
improve understanding of the underlying triggering mechanisms. For example,
spatial power law triggering functions were found more effective than normal
kernels, suggesting aftershocks can be triggered at long ranges, and the rate of
aftershock triggering depends on the magnitude of the mainshock. This has led to
improved earthquake forecasting algorithms based on the ETAS model
\citep{Zhuang:2011pb}. \citet{Harte_2012} explored the effects of model
misspecification and boundary effects on model fits, finding that a good fit for
the background component is also essential, as a poor background fit tends to
bias the model to consider background events as triggered events instead,
overestimating the rate of triggering and the expected number of offspring
events \(m\).

Some research suggests that the parameters of the ETAS model are not spatially
homogeneous, and that a more realistic model would allow the parameters to vary
in space. \citet{Ogata_2003} introduced a method which allows parameters to vary
in space, linearly interpolated between values defined at the corners of a
Delaunay triangulation of the space defined by the earthquake locations. To
ensure spatial smoothness in these values, a smoothness penalty term was added
to the log-likelihood. \citet{Nandan:2017bu} took a similar approach,
partitioning the region \(X\) drawing \(q\) points uniformly at random within
\(X\), obtaining the Voronoi tesselation, and allowing each Voronoi cell to have
a separate set of parameters. No spatial smoothness was imposed, and the number
of points \(q\) was selected via BIC.

Similar concerns apply to temporal nonstationarity. \citet{Kumazawa_2014}
considered two approaches to model changes in parameters over time: a
change-point model, in which parameters are fitted separately to events before
and after a suspected change point, and a continuously varying model in which
several parameters, including the triggering rate, were assumed to be
first-order spline functions in time. Temporal smoothness was enforced with a
penalty term in the log-likelihood, and AIC was used to compare the fits in
series of earthquakes recorded in Japan, finding evidence of nonstationarity in
an earthquake swarm.

\subsection{Crime Forecasting}
\label{crime-forecasting}

After the development of ETAS models, \citet{Mohler:2011ft} drew an analogy
between aftershock models and crime. Criminologists have demonstrated that
near-repeat victimization is common for certain types of crime---for example,
burglars often return to steal from the same area repeatedly
\citep{Short:2009cf,Townsley:2003bu,Bernasco:2015fl}, and some shootings may
cause retaliatory shootings soon after \citep{Ratcliffe:2008hs,Loeffler:2016},
typically within just a few hundred meters. These can be treated as
``aftershocks'' of the original crime.

Similarly, several criminological theories suggest the background rate of crime
can be expected to widely vary by place. Routine activities theory
\citep{Cohen:1979cz} states that criminal acts require three factors to occur
together: likely offenders, suitable targets, and the absence of capable
guardians. These factors vary widely in space depending on socioeconomic
factors, business and residential development, and the activities of police or
other guardians (e.g.\ vigilant neighbors). Rational choice theory
\citep{Clarke:1985gt} considers criminals making rational decisions to commit
offenses based on the risks and rewards they perceive---and the availability of
low-risk high-reward crime varies in space. \citet{Weisburd:2015gu}, using crime
data across several cities, argued for a \textit{law of crime concentration},
stating that a large percentage of crime occurs within just a few percent of
street segments (lengths of road between two intersections) in a given city.
Bolstering this, \citet{Gorr_2014} demonstrated that a policing program based on
both chronic hot spots and temporary flare-ups can be more effective than a
program based on only one or the other.

These theories suggest a model of crime which assumes the conditional intensity
of crime occurrence can be divided into a chronic background portion, which may
vary in space depending on a variety of factors, and a self-exciting portion
which accounts for near-repeats and retaliations \citep{Mohler:2011ft}:
\[
  \lambda(s, t) = \nu(t) \mu(s) + \sum_{i:t_i < t} g(s - s_i, t - t_i),
\]
where \(g\) is a triggering function and \(\nu(t)\) reflects temporal changes
from weather, seasonality, and so on. Initially, \(\nu\), \(\mu\), and \(g\)
were determined nonparametrically following Algorithm~\ref{iterative-fit},
though weighted kernel density estimation was too expensive to perform on the
full dataset of 5,376 residential burglaries, so they modified the algorithm to
subsample the dataset on each iteration. An alternate approach, requiring no
subsampling, would be to use a fast approximate kernel density algorithm to
reduce the computational cost \citep{Gray:2003jh}.

\citet{Mohler:2014iu} introduced a parametric approach intended to simplify
model fitting and also incorporate ``leading indicators''---other crimes or
events which may be predictive of the crime of interest. In a model forecasting
serious violent crime, for example, minor offenses like disorderly conduct and
public drunkenness have proven useful in predictions, since they may reflect
behavior which will escalate into more serious crime \citep{Cohen:2007iw}. The
intensity is simplified to make the background constant in time (\(\nu(t) =
1\)), and to incorporate leading indicators, the background is based on a
weighted Gaussian kernel density estimate, in which \(\nu(t) = 1\) and
\begin{align*}
  \mu(s) = \sum_{i=1}^n \frac{\alpha_{M_i}}{2 \pi \eta^2 T} \exp\left(- \frac{\|s -
  s_i\|^2}{2\eta^2} \right),
\end{align*}
where \(T\) is the length of the time window encompassed by the dataset, \(s_i\)
and \(t_i\) the location and time of crime \(i\), \(M_i\) is a mark giving the
\emph{type} of crime \(i\) (where \(M_i = 1\) by convention for the crime being
predicted), and \(\alpha\) is a vector of weights determining the contribution
of each event type to the background crime rate. The sum is over all crimes,
avoiding the additional computational cost of stochastic declustering. The marks
are treated as unpredictable, and only the ground process is estimated, not the
conditional distribution of marks.

Similarly to \(\mu(s)\), the triggering function \(g\) is a Gaussian in space
with an exponential decay in time:
\[
  g(s, t, M) = \frac{\theta_M}{2\pi \omega \sigma^2} \exp(-t / \omega)
  \exp\left(- \frac{\|s\|^2}{2\sigma^2} \right).
\]
\(\theta\) performs a similar function to \(\alpha\), weighting the contribution
of each type of crime to the conditional intensity. The bandwidth parameters
\(\sigma^2\) and \(\eta^2\) determine the spatial influence of a given crime
type, while \(\omega\) determines how quickly its effect decays in time. In
principle, different spatial and temporal decays could be allowed for each type
of crime, but this would dramatically increase the number of parameters.

\citet{Mohler:2014iu} fit the parameters of this model on a dataset of 78,852
violent crimes occurring in Chicago, Illinois between 2007 and 2012. The crime
of interest was homicide, using robberies, assaults, weapons violations,
batteries, and sexual assaults as leading indicators. The resulting model was
used to identify ``hotspots'': small spatial regions with unusually high rates
of crime. Previous research has suggested that directing police patrols to
hotspots can produce measurable crime reductions, with results varying by the
type of policing intervention employed \citep{Braga:2014by}. To test the
self-exciting model's effectiveness in this role, \citet{Mohler:2014iu} compared
its daily predictions to true historical records of crime, finding that it
outperforms methods that consider only fixed hotspots (equivalent to setting
\(\theta_i = 0\) for all \(i\)) and those that only consider near-repeats
(\(\alpha_i = 0\) for all \(i\)).

\subsection{Epidemic Forecasting}
\label{epidemic-forecasting}

Forecasting of epidemics of disease, such as influenza, typically rely on time
series data of infections or infection indicators (such as physician reports of
influenza-like illness, without laboratory confirmation), and hence often rely
on time series modeling or compartment models, such as the
susceptible--infectious--recovered model \citep{Nsoesie_2013}. This data does
not typically include the location and time of individual infections, instead
containing only aggregate rates over a large area.

When individual-level data is available, however, point processes can model the
clustered nature of infections. Spatial point processes have been widely used
for this purpose \citep[chapter 9]{Diggle:2014}, and when extended to
spatio-temporal analysis, self-exciting point processes are a natural choice,
with excitation representing the transmission of disease. Again following the
ETAS literature, \citet{Meyer:2011ct} introduced a self-exciting spatio-temporal
point process model adapted for predicting the incidence of invasive
meningococcal disease (IMD), a form of meningitis caused by the bacterium
\textit{Neisseria meningitidis,} which can be transmitted between infected
humans and sometimes forms epidemics. Unaffected carriers can retain the
bacterium in their nasopharynx, suggesting that observed cases of IMD can be
divided into ``background'' infections, transmitted from an unobserved carrier
to a susceptible individual, and triggered infections transmitted from this
individual to others.

In a dataset of 636 infections observed in Germany from 2002--2008, each
infection's time, location (by postal code), and finetype (strain) was recorded.
The model includes unique features: rather than empirically estimating the
background function, it is composed of a function of population density and of a
vector of covariates (in this case, the number of influenza cases in each
district of Germany, hypothesized to be linked to IMD). The resulting
conditional intensity function is
\begin{equation*}
  \begin{split}
    \lambda(s, t) ={}& \rho(s, t) \exp\left(\beta' z(s, t)\right)\\
    &{}+ \sum_{j \in I^*(s, t)} e^{\eta_j} g(t - t_j) f(\|s - s_j\|),
  \end{split}
\end{equation*}
where \(I^*(s, t)\) is the set of all previous infections within a known fixed
distance \(\delta\) and time \(\epsilon\). Here \(\rho(s, t)\) represents the
population density, \(z(s, t)\) the vector of spatio-temporal covariates, and
\(\eta_j = \gamma_0 + \gamma' m_j\), where \(m_j\) is a vector of unpredictable
marks on each event, such as the specific strain of infection. The spatial
triggering function \(f\) is a Gaussian kernel, and the temporal triggering
function \(g\) is assumed to be a constant function, as there were comparatively
few direct transmissions of IMD in the dataset from which to estimate a more
flexible function.

The results were promising, showing that the self-exciting model can be used to
estimate the epidemic behavior of IMD. The unpredictable marks \(m_j\) included
patient age and the finetype (strain) of bacterium responsible. Comparisons
between finetypes revealed which has the greatest epidemic potential, and the
age coefficient allowed comparisons of the spread behavior between age groups.

\citet{Meyer:2014in} then proposed to replace \(f\) with a power law function,
previously found to better model the long tails in the movement of people
\citep{Brockmann:2006}. Using the asymptotic covariance estimator given in
eq.~\eqref{sigmahat}, they also produced confidence intervals for their model
parameters, though without verifying the necessary regularity assumptions on the
conditional intensity function \citep[section 4.2.3]{Meyer:2010wz}. A similar
modeling approach was used to test if psychiatric hospital admissions have an
epidemic component, via a permutation test for the parameters of the epidemic
component of the model \citep{Meyer_2016}.

\citet{Schoenberg:2017rj} introduced a recursive self-exciting epidemic model in
which the expected number of offspring \(m\) of an event is not constant but
varies as a function of the conditional intensity, intended to account for the
natural behavior of epidemics: when little of the population has been exposed to
the disease, the rate of infection can be high, but as the disease becomes more
prevalent, more people have already been exposed and active prevention measures
slow its spread. The model takes the form
\[
  \lambda(s, t) = \mu + \int_X \int_0^t H\left(\lambda(s', t')\right) g(s-s',
  t-t') \dif N(s', t'),
\]
where \(g\) is a chosen triggering function and \(H\) is the \emph{productivity
  function}, determining the rate of infection stimulated by each event as a
function of its conditional intensity. \citet{Schoenberg:2017rj} took \(H(x) =
\kappa x^{-\alpha}\), with \(\kappa > 0\), to model decreasing productivity, and
fit to a dataset of measles cases in Los Angeles, California with maximum
likelihood to demonstrate the effectiveness of the model.

\subsection{Events on Social Networks}
\label{social-networks}

The models discussed so far have considered events in two-dimensional space
(e.g. latitude and longitude coordinates of a crime or infection). Recently,
however, self-exciting point processes have been extended to other types of
events, including events taking place on social networks.

\citet{Fox:2016de} considered a network of officers at the West Point Military
Academy. Each officer is a node on the network, and directed edges between
officers represent the volume of email sent between them. \citet{Fox:2016de}
developed several models, the most general of which models the rate at which
officer \(i\) sends email as
\[
  \lambda_i(t) = \nu_i \mu(t) + \sum_j \sum_{r_k^{ij} < t} \theta_{ij} \omega_i
  e^{-\omega_i (t - r_k^{ij})}.
\]
Here \(r_k^{ij}\) represents the time of the \(k\)th message sent from officer
\(j\) to officer \(i\), \(\omega_i\) is a temporal decay effect for officer
\(i\), and \(\theta_{ij}\) models a pairwise reply rate for officer \(i\)'s
replies to officer \(j\). The background rate \(\mu(t)\) is allowed to vary in
time to model time-of-day and weekly effects, with a offset \(\nu_i\) for each
officer. The model is fit by expectation maximization and standard errors found
by parametric bootstrap.

\citet{Zipkin:2015hq} considered the same dataset, but instead of modeling a
self-exciting process for each officer, they assigned one to each edge between
officers, which enabled them to develop methods for a missing-data problem: can
the sender or recipient be inferred if one or both are missing from a given
message? The self-exciting model had promising results, and they suggested a
possible application in inferring participants in gang violence.

Taking an alternate approach, \citet{Green:2017fg} modeled the contagion of gun
violence through social networks in Chicago. The network nodes were all
individuals who had been arrested by Chicago police during the study period,
connected by edges for each pair of individuals who had been arrested together,
assumed to indicate strong pre-existing social ties. Rather than predicting the
rate on edges, as \citet{Fox:2016de} did, this study modeled the probability of
each individual being a victim of a shooting as a function of seasonal
variations (the background) and social contagion of violence, as the probability
of being involved in a shooting is assumed to increase if someone nearby in the
social network was recently involved as well.

This is formalized in the conditional intensity for individual \(k\),
\[
  \lambda_k(t) = \mu(t) + \sum_{t_i < t} \phi_{k_i, k} (t - t_i),
\]
where \(\mu(t)\) represents seasonal variation and the self-excitation function
\(\phi_{k_i,k}\) is composed of two pieces, a temporal decay \(f_\beta(t)\) and
a network distance \(g_\alpha(u, v)\):
\begin{align*}
  f_\beta(t) &= \beta e^{-\beta t}\\
  g_\alpha(u, v) &= \begin{cases}
    \alpha\, \text{dist}(u, v)^{-2} & \text{when } \text{dist}(u, v) \leq 3\\
    0 & \text{otherwise}
  \end{cases}\\
  \phi_{u,v}(t) &= f_\beta(t) g_\alpha(u, v),
\end{align*}
where \(\text{dist}(u, v)\) is the minimum distance (number of edges) between
nodes \(u\) and \(v\). The model was fit numerically via maximum likelihood, and
a form of declustering performed by attributing each occurrence of violence to
the larger of the background \(\mu(t)\) or the sum of contagion from previous
events, rather than using a stochastic declustering method as discussed in
Section~\ref{declustering}.

\section{Conclusions}

When a spatio-temporal point process can be divided into clusters of events
triggered by common causes, self-exciting models are a powerful tool to
understand the dynamics of the process. This review has highlighted developments
in several areas of application which enable fast maximum likelihood and
Bayesian estimation, declustering of events, and a variety of model diagnostics.
Not all of these tools are widely adopted, particularly graphical diagnostics
which have only been developed over the past few years, and there are many open
problems: Bayesian estimation, for example, could lead to hierarchical models
which consider several separate realizations of a process (such as crime data
from different cities), and the application of self-exciting models to data on
networks is in its infancy, and likely has many other possible applications.

Interpretation of self-exciting models does require care, however. For example,
consider an infectious disease with no known carriers---all transmission is from
infected to susceptible individuals, and any given case could, in principle, be
traced back to the index case. The division into background and cluster
processes makes less conceptual sense here, since there is not a background
process producing new cases from nowhere; if a self-exciting model were fit to
infection data, the background process would capture cases caused by unobserved
infections, and an improved rate of case reporting would decrease the apparent
importance of the background process. Unobserved infections would also mean that
\(m\), the estimated number of infections triggered by each case, would be
underestimated, as some triggered infections would not appear in the data.

But when the underlying generative process is clustered, self-exciting
spatio-temporal point processes are fast, flexible, and interpretable tools,
with growing application in many scientific fields.

\section*{Acknowledgments}

The author thanks Joel Greenhouse and Neil Spencer for important suggestions
which improved the manuscript, as well as the referees, Associate Editor, and
discussants for their helpful comments.

This work was supported by Award No.\ 2016-R2-CX-0021, awarded by the National
Institute of Justice, Office of Justice Programs, U.S.\ Department of Justice.
The opinions, findings, and conclusions or recommendations expressed in this
publication are those of the author and do not necessarily reflect those of the
Department of Justice.

\bibliography{review-refs}

\end{document}